\documentclass[pra,twocolumn,floatfix]{revtex4}
                        \usepackage{graphicx}
                        \usepackage{dcolumn}
                        
\begin{document}

                        \def\be{\begin{equation}}
                        \def\ee{\end{equation}}
                        \def\ba{\begin{eqnarray}}
                        \def\ea{\end{eqnarray}}
                        \def\bas{\begin{eqnarray*}}
                        \def\eas{\end{eqnarray*}}

                        %\addtolength{\topmargin}{2cm}

\title{Nonlocal interactions in collapsing condensates in a box}

\author{Stavros Theodorakis and Nikolas Charalambous}
                        \affiliation{Physics Department, University of Cyprus,
P.O. Box 20537, Nicosia 1678, Cyprus}
                        \email{stavrost@ucy.ac.cy}
\date{\today}

\begin{abstract}
The collapse of attractive Bose-Einstein condensates in a box with tunable interatomic interactions was studied experimentally recently. Not only were remarkably stable remnant condensates observed, but furthermore they often seem to involve two stable plateaus. We suggest that these plateaus correspond in fact to two  minima of the energy, the attractive atomic interactions being nonlocal. We show in detail that all the experimental data for these remnant condensates can be accounted for by minimising a variational energy involving a nonlocal interatomic potential that is attractive everywhere and that has a range of the order of the magnetic length.
\end{abstract}

\maketitle

\vskip 0.3cm
\vskip 0.3cm

{\bf I. Introduction}

Detailed experiments on collapsing attractive Bose-Einstein condensates in magnetic traps have revealed the existence of remarkably stable remnant condensates that survived with almost constant number of atoms for more than 1 second\cite{Donley}. In the aftermath of these experiments, many attempts were made to explain the stability of these remnant condensates using the usual local Gross-Pitaevskii equation, augmented by a three body loss term\cite{GP users}:

\be
 \label{GPeqn}
i\hbar\frac{\partial\psi}{\partial t}=-\frac{\hbar^{2}}{2m}\nabla^{2}\psi+\frac{m\omega^{2}r^{2}}{2}\psi+g|\psi|^{2}\psi-i\frac{\hbar K_{3}}{2}|\psi|^{4}\psi,
   \ee
where $\int|\psi|^{2} d^{3}r=N$ and $g=4\pi\hbar^{2}a/m$, $N$ being the particle number in the condensate and $a$ being the scattering length. For a localised condensate wavefunction $\psi$ this equation leads however to the depletion rate

\be
\label{attrition}
\frac{dN}{dt}=-K_{3}\int|\psi|^{6} d^{3}r.
\ee

This equation shows most clearly that the particle number can not remain constant if $\psi$ is nonzero. Thus the survival of a remnant condensate for very long times of the order of one second cannot be explained this way.

A way out is offered when we recall that the Gross-Pitaevskii equation assumes that the attractive interactions between the atoms are very short-ranged and can be approximated by a delta function. It has been shown, however, that if these interactions are nonlocal, then the condensate cannot collapse\cite{Konotop}. The energy possesses  then two minima near the critical point, a local one that corresponds to the usual low density metastable condensate, and an absolute one, that corresponds to a high density stable remnant condensate. The collapse of the metastable condensate and the formation of the remnant condensate is nothing else then than the transition from the local to the absolute minimum\cite{nonlocal}. A recent work\cite{Andreas} used this idea in order to explain in detail the experimental observations of Ref. 1.

This remarkable stability of the remnant condensates for times as long as one second has been studied thoroughly recently for attractive Bose-Einstein $^{39}K$ condensates with tunable interactions in the uniform potential of a cylindrical optical-box trap\cite{Eigen}. In interaction-quench experiments an attractive condensate was prepared above the negative critical scattering length $a_{c}$ and then the scattering length was quenched to a variable scattering length below $a_{c}$, initiating thus the collapse. After a variable holding time the scattering length was turned abruptly from negative to positive and the trap was switched off. Then the cloud was observed.

\begin{figure}[t]
\vskip 0.3cm
                        \includegraphics[width=0.49\textwidth]{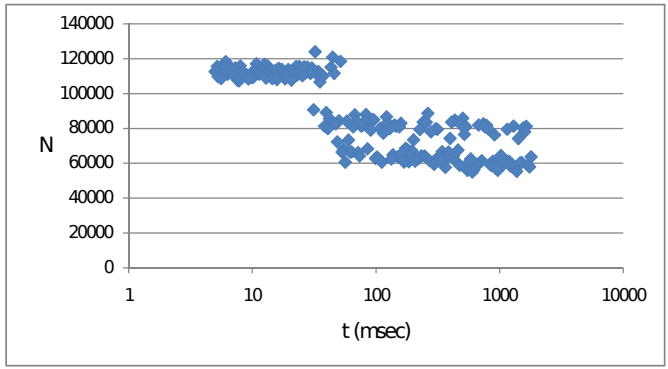}
                        \caption{\label{fig1}Particle number in the condensate  versus time for $a=-1.018a_{0}$ and $L=30.2\mu m$, from set 2 of Ref.\cite{sets}. The maximum number 124021 of the top plateau is the initial number, the top particle number in the intermediate plateau is 90700 and the lowest particle number in the lower plateau is 54913.}
                        \end{figure}

In many instances in these experiments a remarkable two-plateaus structure appeared, as can be seen in the example of Figure~\ref{fig1}.

Indeed, two clearly resolved branches of final particle numbers were observed after the collapse for many of the final values of the scattering length. The final particle numbers did not seem to decline as time passed. Consequently, if one tries to use nonlocal interactions, one would need two minima in the energy after the collapse, not one. Furthermore, these minima should give the particle numbers observed in the experiment of Ref.\cite{Eigen}.

In this paper, we adopt the assumption of nonlocal interatomic interactions in order to interpret these observations. We use a nonlocal interatomic potential derived through general plausible arguments and we then calculate the energy using a very simple variational trial function for the condensate wavefunction. The resulting energy is minimised and the minima are fitted to all six sets of data provided by the experimenters\cite{sets}. We find that the same values for the parameters of the interatomic potential can explain all six sets. Hence the recent experimental data strongly suggest that the nonlocal interatomic interactions are the reason for the appearance of stable remnant condensates manifesting a two-plateaus structure as far as the final particle number is concerned. The condensate makes a transition from a minimum of the initial energy to one of the two available minima of the final energy.

\vskip 0.3cm
\vskip 0.3cm

{II. \bf The nonlocal energy}

The collapse of attractive condensates is prevented by nonlocality. The collapse to a singularity is avoided because the long-range potentials of the interatomic interaction facilitate the formation of big clouds with a rather low density. Thus the question is not whether the initial metastable condensate will collapse into a singularity, but indeed to which minimum of the energy it makes a transition to as it collapses. 

The energy functional for a nonlocally attractive Bose-Einstein condensate in a cylindrical box trap is:
            \ba
            \label{E3D}
           &&E=\int\,\,d^{3}r\,\frac{\hbar^{2}}{2m}|\nabla\Psi|^{2}\nonumber\\
           &&+\frac{g}{2}\int\,\,d^{3}r\,\int\,\,d^{3}r^{\prime}\,|\Psi({\bf r})|^{2} V(|{\bf r}-{\bf r^{\prime}}|)|\Psi({\bf r^{\prime}})|^{2}
            \ea
            where $\int\,d^{3}r|\Psi|^{2}=N$, $N$ being the number of particles. If $a$ is the negative scattering length, then we keep the interatomic potential negative and we set $g=4\pi |a|\hbar^{2}/m$. We note that the particular cylindrical box trap used in Ref.\cite{Eigen} has length $L$ and radius $R=L/2$, $L$ being 30.2 $\mu m$ in most of the cases.

Let us make the energy of Eq.~(\ref{E3D}) dimensionless by measuring distances in units of $R$, energies in units of $\hbar^{2}N/(2mR^{2})$,  the interaction potentials in units of $1/R^{3}$ and wavefunctions in units of $\sqrt{N/R^{3}}$. We obtain

  \ba
            \label{E3Ddimless}
           &&E=\int\,\,d^{3}r\,|\nabla\psi|^{2}\nonumber\\
           &&+\alpha\int\,\,d^{3}r\,\int\,\,d^{3}r^{\prime}\,|\psi({\bf r})|^{2} V(|{\bf r}-{\bf r^{\prime}}|)|\psi({\bf r^{\prime}})|^{2}
            \ea
            where $\alpha=4\pi |a|N/R$, $\int\,d^{3}r|\psi|^{2}=1$ and $\int\,d^{3}r\,V(|{\bf r}|)=-1$. The potential is taken to have an integral of -1, since it will be a generalisation of the negative delta-function potential used in the Gross-Pitaevskii equation.

Since the trap is not a magnetic one, there is no harmonic oscillator term in the energy. The only available length scales are then the scattering length $a$, the radius of the cylinder $R$ and the magnetic length $\sqrt{\hbar/eB}$, which is the smallest size allowed by the uncertainty priciple for a localised orbit within a magnetic field. However, right after collapse a spherical expanding shell was observed in the box, according to Ref.\cite{Eigen}. This shell reflected off the box walls and returned to the remnant. This could only happen if the border of the condensate is at a substantial distance from the walls of the trap. Hence the condensate is considerably smaller than the box and cannot really feel the cylindrical symmetry of the walls. We deduce thus that the condensate wavefunction is spherically symmetric around the centre of the condensate and becomes zero long before reaching the walls. In other words, the length scale $R$ is irrelevant for the condensate.

On the other hand, the scattering length is too small to be an interaction range for the nonlocal interactions. Thus the only length scale that can be associated with this range is the magnetic length, which is equal to $2.56\,\mu m/\sqrt{B}$ if $B$ is measured in Gauss. The scattering length $a$ near a Feshbach resonance is given by the expression

\be
\label{aFeshbach}
a(B)=a_{bg}\big(1-\frac{\Delta}{B-B_{p}}\big),
\ee

where $a_{bg}$ is the background scattering length, $\Delta$ is the resonance width and $B_{p}$ is the resonance centre, these quantities having the values\cite{Derrico} $a_{bg}=-29a_{0}$, $\Delta=-52G$ and $B_{p}=402.45G$ for the $^{39}K$ condensates used in \cite{Eigen}.

We define the positive dimensionless variable $\epsilon=-a/a_{0}$, where $a_{0}$ is the Bohr radius. Then the magnetic length is 

\be
\ell=\frac{2.56556\,\mu m}{\sqrt{402.45+\frac{52}{(\epsilon/29)-1}}}.
\ee

As far as the interatomic potential is concerned, we can assume that it will have a range of the order of $\ell$. The simplest way for the potential to decrease would be a gaussian. We also want it to become zero at some point sufficiently far away from the centre of the condensate, just as the delta-function does in the local case. Furthermore, we expect the interatomic potential to be spherically symmetric, as mentioned earlier. 

We adopt then the simplest spherically symmetric potential that becomes zero at a certain radius, that is long-ranged and that is always attractive and never changes sign: 
\be
\label{V(r)}
V({\bf r})=-\frac{v_{0}}{d^{3}}e^{-r^{2}/d^{2}}\bigl(1-\frac{r^{2}}{q^{2}d^{2}}\bigr)^{2}.
\ee

This potential involves the interaction range $d$ and vanishes at the point $r=qd$, where the parameter $q$ is dimensionless. We demand that its integral over all space be equal to -1, so that it generalises an attractive delta-function. Hence

\be
\label{v0}
v_{0}=\frac{4q^{4}}{\pi^{3/2}(15-12q^{2}+4q^{4})}
\ee

We expect the range $d$ to be of the order of the magnetic length $\ell$, since $a$ is much smaller than $\ell$. So, writing $d$ in units of $R$:

\be
\label{range}
d=c_{0}\frac{\ell}{R},
\ee
where the number $c_{0}$ will be determined from the experimental data. We also expect the parameter $q$ to be a function of the ratio $a/\ell$ of the only two relevant length scales, since it is dimensionless. In any case, its variation with respect to this ratio will have to be a slow one, because this ratio is very small for the magnetic fields used in the experiment.

We shall now adopt the simplest possible variational trial function for the condensate wavefunction:

\be
\label{trial}
\psi_{\gamma}(r)=(\frac{2\gamma}{\pi})^{3/4}e^{-\gamma r^{2}},
\ee

where $\int_{0}^{\infty}4\pi r^{2}|\psi_{\gamma}(r)|^{2}\,dr=1$.

We can calculate then the total energy of  Eq.~(\ref{E3Ddimless}), by integrating over all space. We obtain

  \ba
  \label{E3Dfinal}
 &&E(k)=\frac{3k}{d^{2}}\nonumber\\
&&-\frac{\alpha k^{3/2}(15 - 12(1 + k) q^{2} + 4(1 + k)^{2}q^{4})}{d^{3}\pi^{3/2}(1+k)^{7/2}(15 - 12 q^{2} + 4 q^{4})},
\ea

where the dimensionless parameter $k$ is equal to $\gamma d^{2}$. The main result of this paper is this Eq.~(\ref{E3Dfinal}). 

\begin{figure}[t]
\vskip 0.3cm
                        \includegraphics[width=0.49\textwidth]{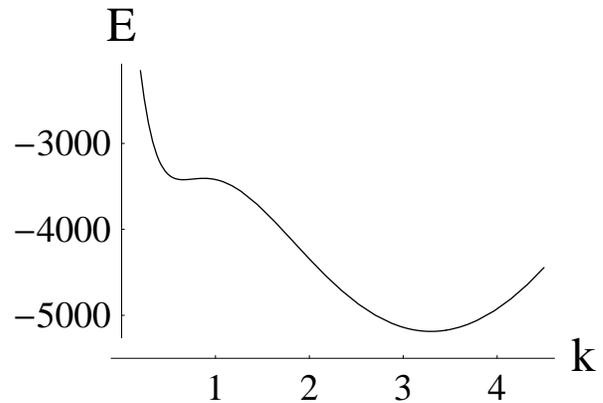}
                        \caption{\label{fig2}The energy of Eq.~(\ref{E3Dfinal}) for the condensate when $q=1.0783$, $d=0.0240$ (in units of $R$) and $\alpha=5.561$.}
                        \end{figure}

We can get an idea of what this energy looks like by plotting it in Figure~\ref{fig2} for the values $q=1.0783$, $d=0.0240$ (in units of $R$), $N=124021$ (a value from set 2 of Ref.\cite{sets}), $a=-1.018a_{0}$, $\epsilon=1.018$, $\alpha=5.561$ and $R=15.1\mu m$.

We can understand now the behaviour of the system when we quench it from a value of $a$ above the critical point. It has two possible minima to make a transition to, hence the two plateaus observed.

If the condensate is in a given initial state $\Psi_{0}$ with $<\Psi_{0}|\Psi_{0}>=N$, we can write it as a superposition of the possible final states
$X_{j}$ through the expansion $\Psi_{0}=\sum_{j}c_{j}X_{j}$, with $<X_{i}|X_{j}>=N\delta_{ij}$, $c_{j}=<X_{j}|\Psi_{0}>/N$ and $\sum_{j}|c_{j}|^{2}=1$. The probability of finding the system in a final state $X_{j}$ is $|c_{j}|^{2}$. In terms of the normalised states $\chi_{j}=\sqrt{R^{3}/N}X_{j}$ and $\psi_{0}=\sqrt{R^{3}/N}\Psi_{0}$ this probability becomes $|<\chi_{j}|\psi_{0}>|^{2}$, leading to the fraction of atoms in the final state $N_{j}/N=|<\chi_{j}|\psi_{0}>|^{2}$.

Our energy has however two states as possible destinations, $\chi_{1}$ and $\chi_{2}$, with corresponding condensate numbers $N_{1}$ and $N_{2}$, where $\chi_{1}$ is taken to have higher energy than $\chi_{2}$. Then a transition from the initial state $\psi_{0}$ can lead either to $\chi_{1}$ or $\chi_{2}$. The transition to $\chi_{2}$ can proceed via two paths: either directly from the initial state to $\chi_{2}$ or  in two stages through the intermediate state $\chi_{1}$ (first from $\psi_{0}$ to  $\chi_{1}$ and then from  $\chi_{1}$ to  $\chi_{2}$). The total probability for this transition to $\chi_{2}$ is the sum of the probabilities for the two paths, leading thus to the condensate number fractions

\be
\label{N1}
\frac{N_{1}}{N}=|<\chi_{1}|\psi_{0}>|^{2}
\ee

and

\be
\label{N2}
\frac{N_{2}}{N}=|<\chi_{2}|\psi_{0}>|^{2}+|<\chi_{2}|\chi_{1}>|^{2}|<\chi_{1}|\psi_{0}>|^{2}.
\ee

Let the wavefunctions $\psi_{0}=\psi_{\gamma_{0}}(r)$, $\chi_{1}=\psi_{\gamma_{1}}(r)$ and $\chi_{2}=\psi_{\gamma_{2}}(r)$ involve the parameters $\gamma_{0}=k_{0}/d^{2}$, $\gamma_{1}=k_{1}/d^{2}$ and $\gamma_{2}=k_{2}/d^{2}$. Then the above equations become:

\be
\label{N1ton}
\frac{N_{1}}{N}=8\frac{(k_{0}k_{1})^{3/2}}{(k_{0}+k_{1})^{3}}
\ee

and

\be
\label{N2ton}
\frac{N_{2}}{N}=8\frac{(k_{0}k_{2})^{3/2}}{(k_{0}+k_{2})^{3}}+64\frac{(k_{2}k_{1})^{3/2}}{(k_{2}+k_{1})^{3}}\frac{(k_{0}k_{1})^{3/2}}{(k_{0}+k_{1})^{3}}.
\ee

These two equations can determine $k_{1}/k_{0}$ and $k_{2}/k_{0}$ if $N_{1}/N$ and $N_{2}/N$ are known. We expect the intermediate state $\chi_{1}$ to have more atoms than the state $\chi_{2}$ and therefore larger size and smaller $k$, since the condensate sheds atoms when going from a higher energy state to a lower energy state. Consequently, we expect $k_{1}<k_{2}$, just as in Figure~\ref{fig2}.

\vskip 0.3cm
\vskip 0.3cm

{III. \bf Determination of the parameters of the potential}

In order to determine the parameter $d$ we shall examine the numbers of atoms in the condensate that were observed in set 2 of Ref.\cite{sets} ($L=30.2\mu m$) when $a$ took three consecutive values, since the variation of the other parameter $q$ is going to be very slow in that case. All the corresponding observed values exhibit the texture of two plateaus for the remnant condensate (see for example Figure~\ref{fig1}, which corresponds to one of the examined values). The initial number $N$ is the highest number of atoms that was observed for the particular value of $a$. The highest number of atoms that was observed below the gap that separates the initial plateau from the intermediate plateau will be called $N_{1}$ and will correspond to the minimum $\chi_{1}$ of the potential, which has the higher energy. The smallest number of the lowest plateau will correspond to the absolute minimum of the potential. Since there is no possible remnant state that has energy below the energy of $\chi_{2}$, the corresponding number of atoms $N_{2}$ will necessarily be the smallest observed number of atoms for the particular value of $a$. The numbers recorded between $N_{1}$ and $N_{2}$ are snapshots of the condensate as it is shedding atoms while passing from the state $\chi_{1}$ to the state $\chi_{2}$.

According to Ref.\cite{sets}, for the value $a=-0.991963a_{0}$ ($\epsilon=0.991963$) we have $N=118320$, $N_{1}=90200$ and $N_{2}=53526$. The corresponding values of $k_{1}/k_{0}$ and $k_{2}/k_{0}$ that are deduced from Eq.~(\ref{N1ton}) and Eq.~(\ref{N2ton}) are 2.3716 and 12.2640 respectively.  For the value $a=-1.01811a_{0}$ ($\epsilon=1.01811$) we have $N=124021$, $N_{1}=90700$, $N_{2}=54913$, $k_{1}/k_{0}=2.5330$ and $k_{2}/k_{0}=12.7009$. For the value $a=-1.04422a_{0}$ $(\epsilon=1.04422)$ we have $N=123219$, $N_{1}=89478$, $N_{2}=55959$, $k_{1}/k_{0}=2.5605$ and $k_{2}/k_{0}=12.4363$. 

Since these values correspond to consecutive values of $a$, we can assume that the parameter $q$ is just a linear function of the small ratio $a/d$ in this region, and hence a linear function of $\epsilon/d$ of the form $p_{0}+p_{1}\epsilon/d$, where $d$ is given by Eq.~(\ref{range}). For each of the three values of $a$ mentioned above there is an unknown $k_{0}$. Along with these three unknown values of $k_{0}$ also unknown are the parameters $c_{0}$, $p_{0}$ and $p_{1}$. We have however six corresponding equations involving these parameters, since for each of these values of $a$ the expression $dE/dk$ vanishes when $k=k_{1}$ and $k=k_{2}$. The solution of these six equations yields the values  0.240094, 0.259505 and  0.271214  for $k_{0}$ when $\epsilon$ becomes 0.991963, 1.01811 and 1.04422,  respectively. We also find that $p_{0}=1.09972$, $p_{1}= -0.000504$ and $c_{0}= 2.63784$.

The value of $c_{0}$ found above can be used for all measured values of $a$, for all sets in Ref.\cite{sets}, since it is just the proportionality constant between $d$ and the magnetic length. Hence the dimensionless $d$ is given by

\be
\label{finald}
d=2.63784\frac{\ell}{R}.
\ee

This interaction range is of the same order of magnitude as the one found in Ref.\cite{Andreas} for condensates in harmonic oscillator traps.

Now that we have found $d$, we can use it for every value of $a$ examined in Ref.\cite{sets} in order to find $q$ as a function of $\epsilon/d$. At the end, we shall compare all these values of $q$ to see if they agree.

First, we need to know $N$, $N_{1}$ and $N_{2}$ for every given value of $a$. The maximum and minimum number of atoms recorded for this value of $a$ are the initial number $N$ and the number of atoms $N_{2}$ at the absolute minimum of the energy, respectively. The number $N_{1}$ is the highest number below $N$ that can allow the existence of two minima in the energy. The numbers of atoms recorded above or below $N_{1}$ are simply snapshots of the condensate as it is shedding atoms during its transition from one state to another. The deduced $N_{1}$ is usually the number of atoms at the top of the intermediate plateau. We begin the determination of the appropriate $N_{1}$ by picking randomly  a value for $N_{1}$ among the values observed, finding the corresponding values of $k_{1}/k_{0}$ and $k_{2}/k_{0}$ and then seeking the values of $q$ and $k_{0}$ for which $E(k_{1})$ and $E(k_{2})$ are minima of the energy. If this value of $N_{1}$ does give minima of the energy, there will always exist another choice for $q$ and $k_{0}$ that also gives two minima. We choose among them the case that gives the lower energy graph. If the picked value of $N_{1}$ works, we then repeat this process with the recorded number of atoms that is exactly above the one that was just examined. This way we find at some point a maximum acceptable value of $N_{1}$, above which the derivative $dE/dk$ cannot be made to vanish at both $k_{1}$ and $k_{2}$. This is the actual value that corresponds to the state $\chi_{1}$. The corresponding values of $q$ and $k_{0}$ are the ones that we are looking for.

We have calculated the values of $N_{1}$ for every value of $a$ that was used in Ref.\cite{sets}. These values, along with the corresponding values for $N$ and $N_{2}$, appear in the appendix.

\begin{figure}[t]
\vskip 0.3cm
                        \includegraphics[width=0.49\textwidth]{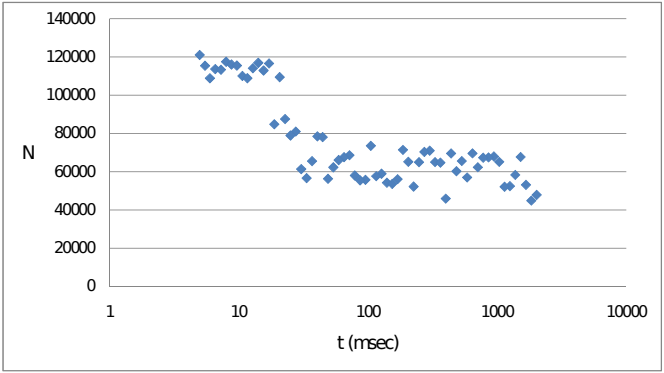}
                        \caption{\label{fig3}Particle number in the condensate  versus time for $a=-1.35369a_{0}$ and $L=30.2\mu m$, from set 2 of Ref.\cite{sets}. The maximum number 121063 of the top plateau is the initial number, the top particle number in the intermediate plateau is 87572 and the lowest particle number in the lower plateau is 44949.}
                        \end{figure}

\begin{figure}[t]
\vskip 0.3cm
                        \includegraphics[width=0.49\textwidth]{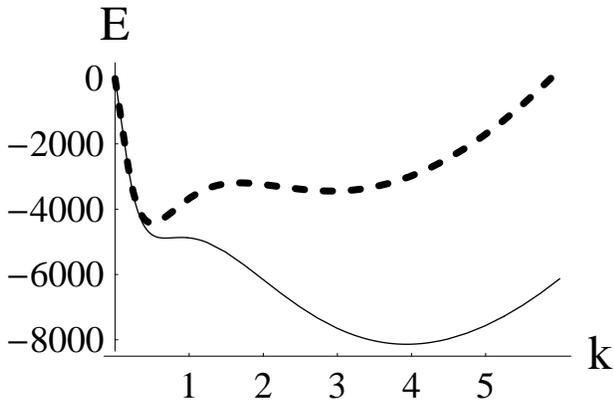}
                        \caption{\label{fig4}The energy E(k) when $\epsilon=1.35369$ and $L=30.2\mu m$, from set 2 of Ref.\cite{sets}. The dashed curve corresponds to $q=1.020045$ and $k_{0}=7.53372$, while the continuous curve corresponds to $q=1.05322$ and $k_{0}=0.263488$. Energetically preferable is the continuous curve, for which the condensate at the absolute minimum has a smaller radius than the condensate at the local minimum.}
                        \end{figure}

As an example, let us apply this method to the examination of the numbers recorded at $\epsilon=1.35369$ in set 2 of Ref.\cite{sets} ($L=30.2\mu m$). These numbers are shown in Figure~\ref{fig3}. The largest among them is $N= 121063$, while the smallest of these is $N_{2}=44949$. If we try and solve the equations $E^{\prime}(k_{1})=E^{\prime}(k_{2})=0$ along with Eq.~(\ref{N1ton}) and Eq.~(\ref{N2ton}) when $N_{1}=108884$, we find that there is no solution. If we examine the next lower number of atoms though, $N_{1}=87572$, we find that there are two solutions. One solution is the configuration $q=1.020045$, $k_{0}=7.53372$, $k_{1}=2.92495$, $k_{2}=0.504465$, $E(k_{1})=-3444.25$, $E(k_{2})=-4424.37$. The second solution is the configuration $q=1.05322$, $k_{0}=0.263488$, $k_{1}=0.678661$, $k_{2}=3.93496$, $E(k_{1})=-4885.16$, $E(k_{2})=-8136.43$. This second configuration is energetically preferable, as shown in Figure~\ref{fig4}. Hence the radius of the condensate is smallest when it has reached the absolute minimum. In fact, we have verified that the local minimum is always on the left in the energy $E(k)$, for all values of $a$ and for all the sets. Since the value $N_{1}=87572$ is the highest for which there exists a configuration for $q$ and $k_{0}$ with two minima, it is the appropriate number of atoms for the condensate when it is in the local minimum of the energy for the case $\epsilon=1.35369$. Lower numbers of atoms correspond to snapshots of the condensate during the transition from the local to the absolute minimum of the energy.

Using the method decsribed above we have calculated the values of the parameter $q$ for every value of $a$ that was used in Ref.\cite{sets}. Furthermore, we have found the critical point for each set, since at the critical point the energies of the two minima are equal. The value of $\alpha$ at the critical point is 4.080913 for all the sets. The values at the critical point for $q$, $\epsilon$ and $N$ for each set are given in Table I.

\begin{table}[t]
\vskip 0.3cm
                        \includegraphics[width=0.49\textwidth]{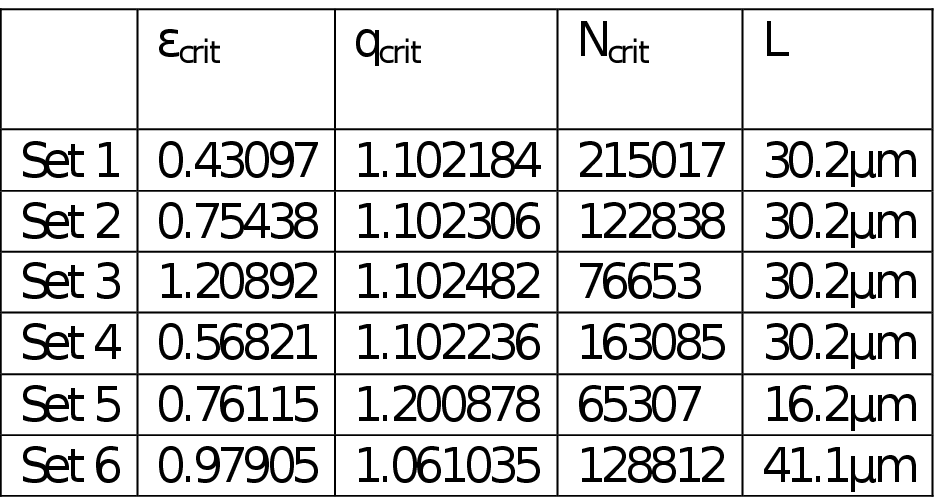}
                        \caption{\label{Table I}The values at the critical point for $q$, $\epsilon$ and $N$ for each set. The corresponding value of $\alpha$ is 4.080913 for all the sets.}
                        \end{table}

\begin{figure}[t]
\vskip 0.3cm
                        \includegraphics[width=0.49\textwidth]{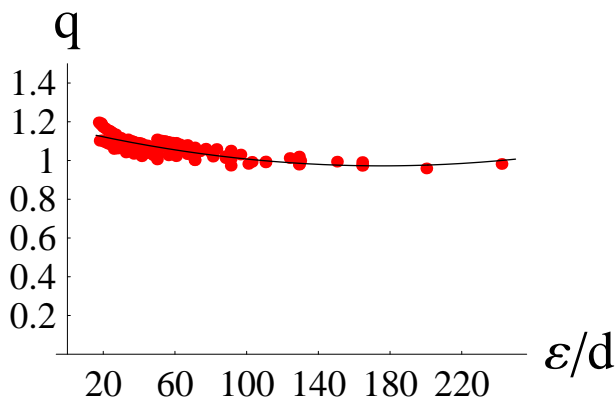}
                        \caption{\label{fig5}The values of $q$ as a function of $\epsilon/d$ for all six sets of data of Ref.\cite{sets}. A parabolic fit to these data is the function $1.16335-0.00217809\epsilon/d+6.2119\times 10^{-6}\epsilon^{2}/d^{2}$.}
                        \end{figure}

Having found the value of $q$ for each value of $a$ used, we plot them in Figure~\ref{fig5} as a function of $\epsilon/d$ for all six sets of Ref.\cite{sets}. We can see that for any given $a$ the values of $q$ roughly coincide, irrespective of the set they belong to and irrespective of $N$. This confirms the fact that $q$ is a parameter of the interatomic potential and must be the same function of $\epsilon/d$ for all values of $N$ and all the sets.

\begin{figure}[t]
\vskip 0.3cm
                        \includegraphics[width=0.49\textwidth]{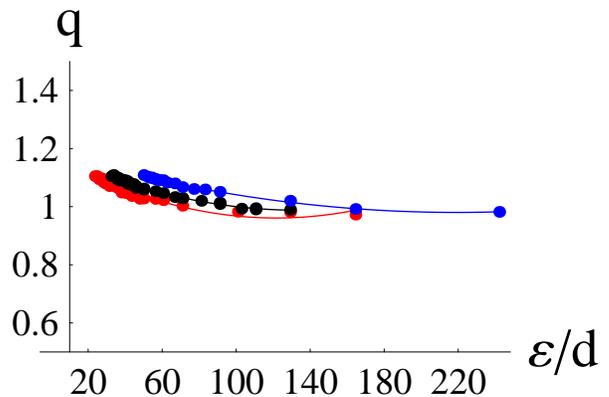}
                        \caption{\label{fig6}The values of $q$ as a function of $\epsilon/d$ and their parabolic fits for the sets 2 (black), 3 (blue) and 4 (red) of Ref.\cite{sets}.}
                        \end{figure}

Furthermore, we see that $q$ varies very slowly as a function of $\epsilon/d$. We can assume then that $q$ is a parabolic function of this parameter. We find that all the data are fitted very well by the function

\be
\label{fit}
q=1.16335-0.00217809\epsilon/d+6.2119\times 10^{-6}\epsilon^{2}/d^{2},
\ee

as we can see in Figure~\ref{fig5}. We can also find for each set a parabolic fit for the $q$ values, required to pass through the critical point. These fits are given in Table II. They are good to within 0.5 percent, as can be seen in Table II, as well as in Figure~\ref{fig6} for the case of sets 2, 3, 4.

\begin{table}[t]
\vskip 0.3cm
                        \includegraphics[width=0.49\textwidth]{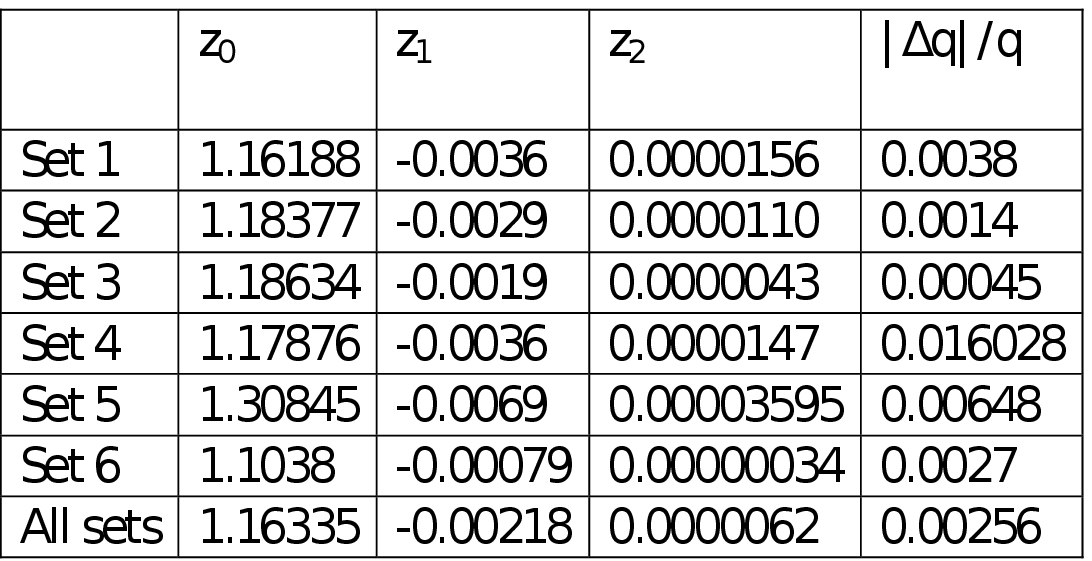}
                        \caption{\label{Table II} The parabolic fits to $q$ in the form $q_{fit}=z_{0}+z_{1}\epsilon/d+z_{2}\epsilon^{2}/d^{2}$ for each set of Ref.\cite{sets}, along with the percentage difference $\Delta q/q=|q_{fit}-q|/q$ of the $q$ values from these fits.}
                        \end{table}

\begin{figure}[t]
\vskip 0.3cm
                        \includegraphics[width=0.49\textwidth]{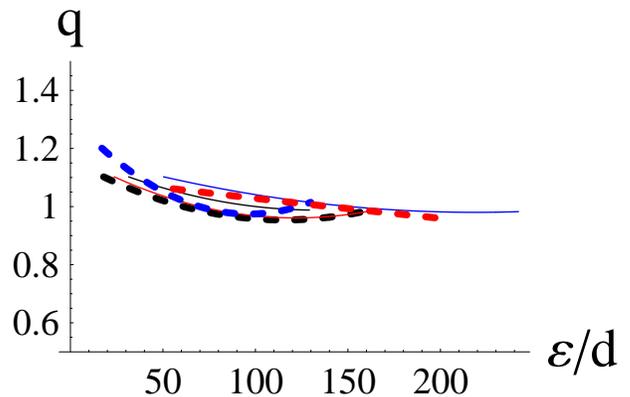}
                        \caption{\label{fig7}The parabolic fits of $q$ as a function of $\epsilon/d$ for set 1 (black dashed), set 2 (black), set 3 (blue), set4 (red), set 5 (blue dashed) and set 6 (red dashed) of Ref.\cite{sets}.}
                        \end{figure}

 In Figure~\ref{fig7} we plot these parabolic fits for the sets. We see that they almost coincide, irrespective of the particular set.

Thus we have found $q$ as a parabolic function of $\epsilon/d$, for each one of the six sets. For each value of $a$ in a certain set though, we have also obtained previously the value of $k_{0}$, the parameter of the initial state of the condensate before the quench. We can then use the equation $E^{\prime}(k_{0})=0$ for the initial state, with $q$ given by the parabolic fit for this particular set, in order to find the parameter $\epsilon_{0}$ and the corresponding scattering length $-\epsilon_{0}a_{0}$ of the initial state (before the quench). We have calculated in fact all these initial scattering lengths and we verified that all of these corresponded to minima of the energy. 

\begin{figure}[t]
\vskip 0.3cm
                        \includegraphics[width=0.49\textwidth]{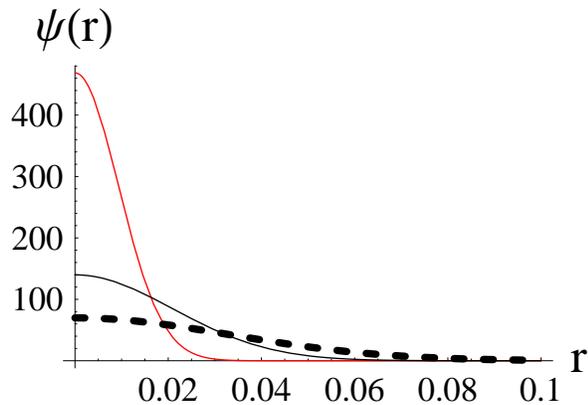}
                        \caption{\label{fig8}The wavefunctions for the initial (dashed line), intermediate (black line) and final (red line) states for the case of $a=-1.01811a_{0}$ from set 2 of Ref.\cite{sets}. Here $\epsilon=1.01811$, $N=124021$, $N_{1}=90700$, $N_{2}=54913$, $q=1.07834$, $q_{0}=1.13879$, $k_{0}=0.259505$, $k_{1}=0.657325$, $k_{2}=3.29594$ and $\epsilon_{0}=0.390966$.}
                        \end{figure}

For example, for the case $a=-1.01811a_{0}$ of set 2 of Ref.\cite{sets} ($\epsilon=1.01811$) we have $N=124021$, $N_{1}=90700$, $N_{2}=54913$, $q=1.07834$, $k_{0}=0.259505$, $k_{1}=0.657325$, $k_{2}=3.29594$ and $\epsilon_{0}=0.390966$. We use the expression of Table II for set 2 in order to find the initial value of $q$, $q_{0}=1.13879$. In Figure~\ref{fig8} we plot the corresponding wavefunctions for the initial, intermediate and final state.

\begin{figure}[t]
\vskip 0.3cm
                        \includegraphics[width=0.49\textwidth]{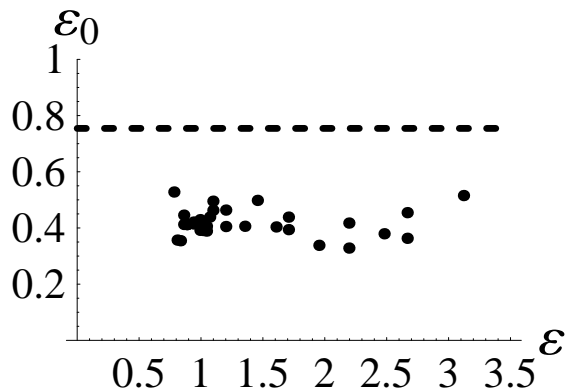}
                        \caption{\label{fig9} The starting points $\epsilon_{0}$ versus the parameter $\epsilon=-a/a_{0}$ for the data of set2 of Ref.\cite{sets}. They are all below the critical value $\epsilon_{crit}=0.754383$.}
                        \end{figure}

We note further that the initial scattering lengths are above the critical scattering length. We show for example in Figure~\ref{fig9} the starting points $\epsilon_{0}$ versus the parameter $\epsilon=-a/a_{0}$ for the data of set2. They are all below the critical value $\epsilon_{crit}=0.754383$, as they should.

\begin{figure}[t]
\vskip 0.3cm
                        \includegraphics[width=0.49\textwidth]{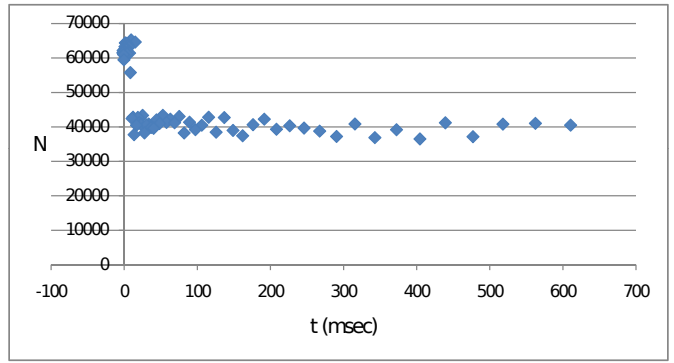}
                        \caption{\label{fig10}Particle number in the condensate  versus time for $a=-1.04422a_{0}$ and $L=16.2\mu m$, from set 5 of Ref.\cite{sets}. The maximum number 65259 of the top plateau is the initial number, the particle number for the higher energy minimum is the top particle number 43358 of the lowest plateau and the lowest particle number 36508 in the lowest plateau is the particle number for the absolute minimum of the energy.}
                        \end{figure}

\begin{figure}[t]
\vskip 0.3cm
                        \includegraphics[width=0.49\textwidth]{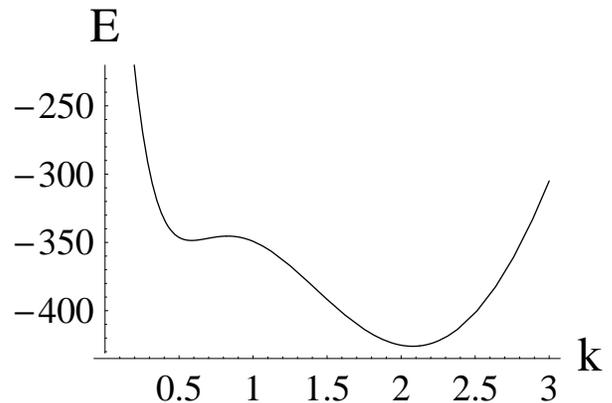}
                        \caption{\label{fig11}The energy of Eq.~(\ref{E3Dfinal}) for the condensate when $q=1.15366$, $d=0.0448$ (in units of $R$), $\alpha=5.594$, $a=-1.04422a_{0}$ and $L=16.2\mu m$, from set 5 of Ref.\cite{sets}.}
                        \end{figure}

One last comment concerns the appearance of the plateaus for a number of the recorded values $a$. When $N_{1}$ is close to $N$ or $N_{2}$, the intermediate plateau coalesces with the initial or the lowest plateau and one gets the impression of a single plateau of collapse. Such is the case with the value $a=-1.04422a_{0}$ from the set 5 of Ref.\cite{sets}. The observed values for the number of atoms are shown in Figure~\ref{fig10},  for which $N=65259$, $N_{1}=43358$ and $N_{2}=36508$. These numbers are the maximum number of the top plateau, the top particle number in the lowest plateau and the lowest particle number in the lowest plateau, respectively. The value 43358 for $N_{1}$ is again the highest one that allows two minima in the energy. We find for this case that $k_{0}=0.200980$, $k_{1}=0.584803$, $k_{2}=2.07838$, $q=1.15366$, $\epsilon_{0}=0.533563$. The corresponding energy is shown in Figure~\ref{fig11}.

{IV. \bf Conclusions}

The observed remarkable stability of the remnant condensate that was observed during the collapse of an attractive Bose-Einstein condensate in a box after quenching from a value of $a$ above the critical point cannot be explained in terms of an extension of the Gross-Pitaevskii equation via an imaginary quintic term. A nonlocal extension of this equation is required. The nonlocal potential will have to be spherically symmetric, since the observations show that the condensate is much smaller than the box and does not feel the cylindrical features of the box, and will have to vanish at some point. Furthermore it should never change sign. As a result of these requirements, the energy will have two minima. This is why two plateaus are observed in the experiment. As the scattering length is changed suddenly to a value below the critical point, the condensate makes a transition to either of the two existing minima. In calculating the fraction of the number of atoms that remains in the remnant condensate we must take into account the fact that there are two possible paths towards the absolute minimum, a direct one and one through the local minimum. We have used a simple trial function in order to find the minima of the energy and we have determined the values of the parameters $q$ and $d$ that give the observed fractions of atoms. The interaction range of the potential is of the order of the magnetic length, while the parameter $q$ is a slow function of $a/d$. All the experimental sets give approximately the same values for $q$ as a function of $a/d$.

{V. \bf Appendix}

The maximum and minimum values observed for the number of atoms at each particular value of $a$ are $N$ and $N_{2}$. The number $N_{1}$ is the maximum observed number for which an energy configuration with two minima is possible at a particular value of $\epsilon=-a/a_{0}$.
The values $(\epsilon, N, N_{1}, N_{2})$ are given below for each set of Ref.\cite{sets}.

Set 1

(0.485633, 219330, 128345, 107227),  (0.512738, 213150, 129616, 111980),  (0.539791, 213760, 134159, 113614),  (0.566793, 208532, 140814, 119245),  (0.593744, 207740, 187947, 122090),  (0.620644, 220370, 212273, 133302),  (0.674291, 209680, 191293, 113109),  (0.780979, 210246, 203164, 114184),  (0.886867, 216956, 208130, 107826),  (0.991963, 225317, 217964, 103667),  (1.19981, 209675, 207777, 94383),  (1.70619, 202270, 176654, 64916),  (2.19445, 209301, 196267, 53167),  (3.12038, 175232, 161432, 43685),  (3.98447, 147903, 134211, 34070)

Set2

(0.780979, 119910, 112301, 86409), (0.807526, 118598, 75541, 52137), (0.834023, 118493, 78930, 51286), (0.860469, 122012, 87377, 64678), (0.860469, 124278, 106385, 72130), (0.886867, 123539, 100002, 64832), (0.886867, 121152, 90447, 62161), (0.939513, 120289, 93051, 60528), (0.939513, 114852, 87646, 55868), (0.965762, 117038, 86403, 56231), (0.991963, 118320, 90200, 53526), (0.991963, 119196, 86449, 58922), (1.01811, 124021, 90700, 54913), (1.04422, 123219, 89478, 55959), (1.04422, 116856, 87217, 50051), (1.07027, 119359, 92156, 56926), (1.09628, 118836, 107045, 64386), (1.09628, 119558, 100270, 59415), (1.19981, 118187, 98348, 54938), (1.19981, 120276, 93232, 49236), (1.35369, 121063, 87572, 44949), (1.45533, 119606, 100627, 51597), (1.6064, 119438, 89704, 38969), (1.70619, 121433, 97230, 41764), (1.70619, 119410, 86310, 35946), (1.95253, 115237, 62826, 22807), (2.19445, 118810, 61593, 20842), (2.19445, 115268, 85793, 30535), (2.47912, 117015, 82009, 25641), (2.66555, 116679, 93289, 29595), (2.66555, 124828, 87084, 25099), (3.12038, 118117, 99841, 29821)

Set 3

(1.19981, 76653, 67938, 50746), (1.25130, 76907, 51725, 33675), (1.27697, 77256, 54997, 35717), (1.30259, 76861, 54349, 35524), (1.32816, 77349, 49562, 30912), (1.35369, 79838, 53586, 35049), (1.37917, 79228, 52994, 32686), (1.40460, 78787, 56010, 35903), (1.45533, 78140, 54798, 37240), (1.50587, 75181, 57136, 34672), (1.60640, 77335, 55555, 33727), (1.70619, 79038, 48061, 25127), (1.85453, 76314, 55552, 29073), (2.00126, 78355, 46199, 25003), (2.19445, 75772, 44453, 21142), (3.12038, 75685, 50359, 18975), (3.98447, 76573, 59983, 18929), (5.91171, 72718, 63169, 17321)

Set 4

(0.566793, 170987, 169853, 146511), (0.593744, 170392, 102014, 88321), (0.593744, 163085, 162607, 142265), (0.620644, 170037, 135641, 89283), (0.647493, 164414, 109727, 90858), (0.674291, 171825, 139223, 88433), (0.701039, 168314, 148203, 94692), (0.727736, 166512, 138505, 89835), (0.754383, 168606, 159911, 103859), (0.780979, 160850, 116903, 85548), (0.834023, 169946, 121306, 79737), (0.860469, 169571, 122110, 84876), (0.886867, 163571, 149270, 87743), (0.913215, 172230, 162887, 91872), (0.965762, 159667, 152785, 87946), (0.991963, 174694, 152705, 79142), (0.991963, 161232, 118327, 72293), (1.01811, 161589, 118749, 74940), (1.04422, 171514, 158644, 80244), (1.14814, 172833, 155952, 71912), (1.19981, 167198, 147991, 67232), (1.35369, 167992, 138825, 61304), (1.45533, 162127, 135974, 57161), (1.70619, 172762, 160776, 60467), (2.43209, 169425, 152478, 43610), (3.12038, 161246, 135152, 35113), (3.98447, 142592, 122699, 28113)

Set 5

(0.780979, 66141, 35252, 33297), (0.807526, 66254, 51679, 51679), (0.834023, 65058, 34672, 30967), (0.834023, 64216, 27631, 27631), (0.834023, 65268, 33993, 30803), (0.860469, 65694, 35292, 30795), (0.886867, 66387, 35221, 29845), (0.939513, 65642, 40916, 34717), (0.991963, 65618, 37956, 30660), (1.04422, 65259, 43358, 36508), (1.09628, 65173, 44348, 35275), (1.09628, 66158, 45204, 37782), (1.14814, 66871, 46099, 36792), (1.19981, 63879, 45297, 34952), (1.19981, 64648, 48650, 37357), (1.32816, 65160, 44985, 31601), (1.45533, 67911, 50018, 33163), (1.70619, 64217, 45041, 27111), (2.66555, 65133, 45670, 22051), (3.98447, 62386, 50282, 18896), (5.91171, 59977, 45775, 14570)

Set 6

(0.991963, 128812, 91260, 62422), (1.04422, 130403, 92494, 59563), (1.09628, 129911, 118577, 68604), (1.14814, 128082, 114290, 65207), (1.19981, 129765, 100561, 51768), (1.25130, 129084, 100685, 53402), (1.30259, 132346, 99002, 48393), (1.40460, 130665, 82563, 40547), (1.55623, 129507, 121325, 56993), (1.70619, 127681, 98129, 44065), (2.19445, 125388, 98593, 36960), (2.66555, 129162, 100377, 30265), (3.55976, 135683, 120154, 29006)

\end{document}